\documentstyle[aps,prb,multicol,graphics]{revtex}
\begin{document}
\draft
\title{Finite-temperature resistive transition in the two-dimensional $XY$ gauge glass model}
\author {Beom Jun Kim}
\address {Department of Theoretical Physics,
Ume{\aa} University,
901 87 Ume{\aa}, Sweden}
\preprint{\today}
\maketitle
\begin{abstract}

We investigate numerically the resistive transition in the two-dimensional $XY$
gauge glass model. The resistively-shunted junction dynamics subject   to the
fluctuating twist boundary condition is used and the linear resistances in the
absence of an external current at various system sizes are computed.  Through
the use of the standard finite-size scaling method, the finite temperature
resistive transition is found at $k_BT_c = 0.22(2)$ (in units of the Josephson
coupling strength) with dynamic critical exponent $z = 2.0(1)$ and the static
exponent $\nu = 1.2(2)$, in contrast to widely believed expectation of the
zero-temperature transition.  Comparisons with existing experiments and
simulations are also made.

\end{abstract}

%75.10.Nr Spin-glass and other random models
%64.70.Pf Glass transitions
%74.50.+r Proximity effects, weak links, tunneling phenomena, and Josephson
%effects
%74.25.Fy Transport properties (electric and thermal conductivity,
%thermoelectric effects, etc.)

\begin{multicols}{2}

Since the prediction of the vortex glass (VG) phase in the high-$T_c$
materials,~\cite{mpafisher89,ffh} the properties and the existence of VG phase
have drawn intensive interest for a
decade.~\cite{YBCO,YBCO:2D,sawa,BiSCCO,yamasaki,cieplak91,Defect,maucourt:moore,cieplak92,huse:reger,apyoung,3DT,mpafisher91,mychoi,reger93,granato98,hyman,yhli,bjkim97,gsjeon,enomoto}
In experiments~\cite{YBCO,YBCO:2D,sawa,BiSCCO,yamasaki} the VG phase has a significant practical importance since the
high-$T_c$ materials can be truly superconducting in this phase where vortices
are completely frozen at random
positions,
while the Abrikosov vortex lattice in pure type-II superconductors dissipate
energy at any amount of external currents.  To study VG from a theoretical point
of view simplified discretized models such as the $XY$ gauge glass and the $XY$
spin glass have been widely used.  For the gauge glass model, there is a
growing consensus that the low critical dimension is less than 3,
which means that there exists VG phase at finite temperatures in three
dimensions (3D).  This has been
confirmed by the defect wall energy calculations at zero
temperature~\cite{cieplak91,Defect,maucourt:moore,cieplak92}
as well as by finite-temperature simulations.~\cite{huse:reger,apyoung,3DT}  Although the
issue about the validity of using the gauge glass model to describe real bulk
high-$T_c$ materials is not completely settled (e.g., the former is isotropic
with a vanishing net external magnetic field, while the latter is anisotropic
with nonzero net magnetic field), experiments on high-$T_c$ materials like
YBCO~\cite{YBCO,YBCO:2D,sawa} and
BiSCCO~\cite{BiSCCO,yamasaki} also have yielded results in accordance with the
existence of a VG phase at finite temperatures. 

In 2D, all existing defect wall energy calculations have unanimously revealed
that the stiffness exponent  has a negative
value,~\cite{Defect,cieplak92,mpafisher91}
which has been confirmed in Ref.~\onlinecite{mychoi} through a simple argument
but has been interpreted as an artifact of ubiquitous spin wave fluctuations.
Finite-temperature simulations have so far yielded contradicting results: On
the one hand, zero-temperature transition has been concluded from Monte-Carlo
(MC) simulations in earlier studies~\cite{apyoung,mpafisher91,reger93} and from
the current-voltage ($IV$) characteristics in resistively-shunted junction
(RSJ) simulations.~\cite{granato98,hyman}  On the other hand, there have been
evidences of the finite-temperature transition from the $IV$
characteristics~\cite{yhli} and from the divergence of the relaxation time
scale~\cite{bjkim97} in RSJ dynamic simulations, and from the very recent
extensive finite-temperature MC simulations.~\cite{mychoi}  Also,
Langevin dynamic simulation of a vortex model with quenched impurities, which
is closely related with the gauge glass model, also has found
the finite-temperature transition in 2D.~\cite{enomoto}  To make things more
complicated, absence of finite-temperature VG phase has been reported in
early experiments on very thin YBCO films,~\cite{YBCO:2D} whereas a
recent study on thin YBCO films, where the correlation length in $c$-direction
exceeds the film thickness,~\cite{sawa} as well as experiments on highly
isotropic BiSCCO [Bi(2:2:2:3)]~\cite{yamasaki} has obtained results which can
be interpreted as indications of finite-$T$ VG phase in 2D.  Consequently, we
believe that the question about the existence of the finite-$T$ VG transition
is not resolved yet. To our knowledge, a detailed numerical study of transport
properties of 2D gauge glass model has not been performed: Existing RSJ
studies~\cite{granato98,hyman,bjkim97} did not  consider the finite-size
scaling in a proper way and in some of them~\cite{granato98,hyman} the
temperature range used did not cover the expected transition temperature ($T_c
\approx 0.22$ in Refs.~\onlinecite{mychoi} and \onlinecite{bjkim97}, and $T_c \approx 0.15$ in
Ref.~\onlinecite{yhli}).

In this paper, we use RSJ dynamic simulation to investigate the resistive
transition in the 2D $XY$ gauge glass model in the absence of external currents.
Since the voltage is always nonzero in the presence of a finite current because
of the nucleation process, vanishing of the {\it linear} resistance, which is
defined in the limit of zero external current, is the appropriate definition of
a superconductor. The fluctuating twist boundary condition (FTBC) for
Langevin-type dynamic simulation~\cite{bjkim99} made it
possible to calculate the linear resistance with  preserved
periodicity of the phase variables (similar methods have also been
used in Ref.~\onlinecite{eikman}).  We here use FTBC to calculate the
linear resistances, and then, through the standard finite-size scaling analysis
(see, e.g., Refs.~\onlinecite{ffh} and \onlinecite{weber}), a resistive transition at a finite
temperature is concluded. 

The Hamiltonian of the 2D $L\times L$ $XY$ gauge glass model 
subject to FTBC is written as~\cite{bjkim99,olsson,lars}
\begin{equation} \label{eq_H}
H = -J\sum_{\langle ij \rangle} \cos(\theta_i - \theta_j 
   - {\bf r}_{ij} \cdot {\bf \Delta}  - A_{ij}), 
\end{equation}
where $J$ is the Josephson coupling strength, the summation is over all
nearest neighboring bonds, and $\theta_i$ 
(with periodicity $\theta_i = \theta_{i + L{\bf \hat x}} = \theta_{i + L{\bf \hat y}}$)
is the phase of the superconducting order parameter at site $i$. 
The twist variable ${\bf \Delta}=(\Delta_x,\Delta_y)$ measures the global twist 
of phase variables in each direction, and  
${\bf r}_{ij} (= {\bf \hat x}, {\bf \hat  y}) $ is the unit vector from site $i$ to the nearest-neighbor
site $j$ (the lattice spacing is taken to be unity).
The $XY$ gauge glass model is characterized by
the quenched random variable $A_{ij}$ uniformly distributed in $[-\pi, \pi)$,
whereas in the $XY$ spin glass model $A_{ij}$ can have binary values $0$ and
$\pi$ with equal probability. Recently $XY$ spin glass model draws attention
in relation to the $\pi$-junctions due to the $d$-wave symmetry of order parameters.

The equations of motion for RSJ dynamics subject to FTBC are determined
from the local and the global current conservations and
are written in dimensionless forms (see Refs.~\onlinecite{bjkim99} and \onlinecite{lars} for details):
For phase variables 
\begin{equation}\label{eq_rsj}
\dot\theta_i = -\sum_j G_{ij}{\sum_k}^{'} [\sin(\theta_j -
\theta_k-{\bf r}_{jk}\cdot {\bf \Delta} -A_{jk}) + \eta_{jk} ], 
\end{equation}
where $G_{ij}$ is the 2D square lattice 
Green's function and the primed summation is over four nearest
neighbors $k$ of site $j$, while for the twist variables 
\begin{equation} \label{eq_dot_delta}
\dot{\bf \Delta } = -\frac{1}{L^2}\frac{\partial H}{\partial {\bf \Delta}} + \eta_{\bf \Delta}. 
\end{equation}
The thermal noise terms satisfy 
$\langle \eta_{ij} \rangle = \langle \eta_{\bf \Delta} \rangle = 
\langle \eta_{ij} \eta_{\bf \Delta} \rangle =  
\langle \eta_{\Delta_x} \eta_{\Delta_y} \rangle = 0$ and
\begin{eqnarray*}
& &\langle \eta_{ij}(t) \eta_{kl}(0) \rangle = 
2T\delta(t)(\delta_{ik}\delta_{jl}-\delta_{il}\delta_{jk}), \\
& &\langle \eta_{\Delta_x}(t) \eta_{\Delta_x}(0) \rangle = 
\langle \eta_{\Delta_y}(t) \eta_{\Delta_y}(0) \rangle = 
\frac{2T}{L^2}\delta(t),
\end{eqnarray*}
where $\langle \cdots \rangle$ is the thermal average, 
and the time $t$ and the temperature $T$ have been normalized in 
units of $\hbar^2/4e^2R_0J$ with the shunt resistance $R_0$,
and $J/k_B$, respectively.
The equations of motion were integrated numerically by using the efficient
second-order Runge-Kutta-Helfand-Greenside algorithm~\cite{batrouni85}
with the discrete time step $\Delta t = 0.05$. For each disorder realization
we neglected an initial stage of the time evolution up to $t_0$, and 
the measurements were made between $t_0$ and $t_0 + t_m$. 
For $L=4$, 6, 8, and 10, disorder averages were performed over 
$N_s = 400$, 200, 100, 50 samples, respectively.
We verified that $t_0 = 10^6$ and $t_m = 5\times 10^6$ (corresponding to $10^8$ time steps)
were sufficiently large; it is more important to increase the 
number of sample averages
to have more reliable results than increasing $t_m$ further.
The highest temperature $T=0.30$ in the present work corresponds to the
lowest temperature in Ref.~\onlinecite{granato98}, and this 
makes it necessary  to use much larger value of $t_m$ than
in Ref.~\onlinecite{granato98} [by factor $O(10^2)$].

The linear resistance in $x$ direction 
is given by the Nyquist formula:~\cite{reif}
\begin{equation}
R_x = \frac{1}{2T} \left[ \int_{-\infty}^\infty dt \langle V_x(t) V_x(0) \rangle \right ]\label{eq:R1}
\end{equation}
with the disorder average $[ \cdots ]$ and 
the voltage across the whole sample $V_x = -L\Delta_x$,
which is then approximated as~\cite{foot:R}
\begin{equation}
R_x \approx \frac{L^2}{2 T } \frac{1}{\Theta} \left[ \left \langle [
\Delta_x(\Theta) - \Delta_x(0) ]^2 \right\rangle \right] , \label{eq:Rmu}
\end{equation}
for sufficiently large $\Theta$. For convenience, we define 
\begin{equation} \label{eq:f}
f(\Theta) \equiv \frac{L^2}{2T} \frac{  
\left[ \left \langle [ \Delta_x(\Theta) - \Delta_x(0) ]^2 \right\rangle \right]  +
\left[ \left \langle [ \Delta_y(\Theta) - \Delta_y(0) ]^2 \right\rangle \right]  
} {2} 
\end{equation}
and calculate the linear resistance $R$ from the least-square fit
to the form $ f(\Theta) = R \Theta$.

\begin{figure}
\centering{\resizebox*{!}{6.0cm}{\includegraphics{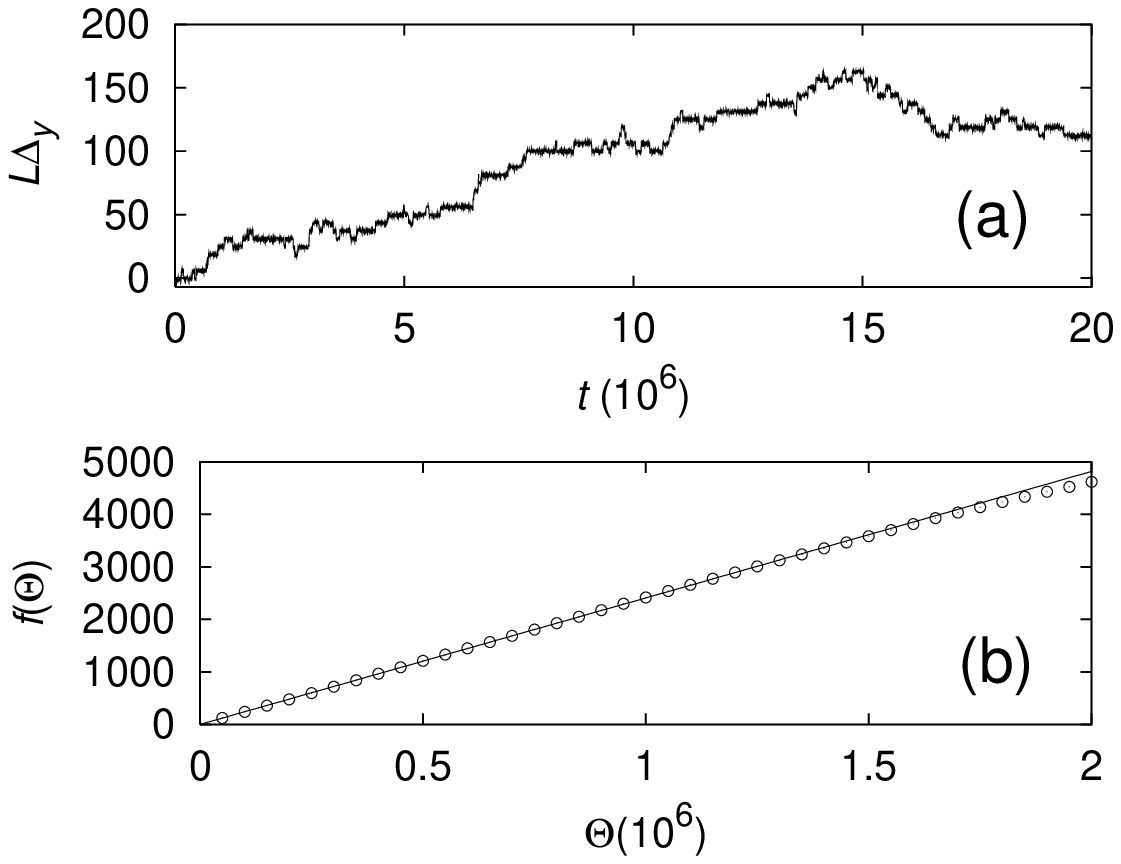}}}
\caption{(a) Time evolution of $L\Delta_y(t)$
for a given disorder realization for $L=6$ at $T=0.22$.
(b) $f(\Theta)$ versus $\Theta$ [see Eq.~(\protect\ref{eq:f})] obtained
from 200 sample averages for $L=6$ at $T=0.22$ (open circles). Least-square fit
to the form $f(\Theta) = R \Theta$ (full line)
determines the linear resistance.
}
\label{fig:delta}
\end{figure}

\begin{figure}
\centering{\resizebox*{!}{6.0cm}{\includegraphics{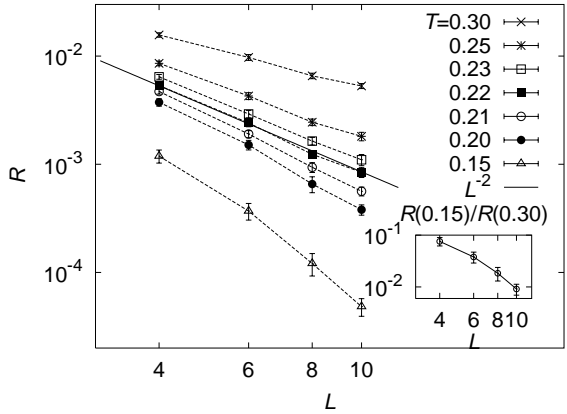}}}
\caption{Linear resistance $R$ versus system size $L$ 
at temperatures $T=0.30$, 0.25, 0.23, 0.22, 0.21, 0.20,
and 0.15 (from top to bottom). It is shown that the behavior
in the high-temperature regime ($T \protect\gtrsim 0.25$) is different
from that in the low-temperature regime ($T \protect\lesssim 0.15$), implying
the existence of the resistive transition near $T = 0.22$, where the power-law
behavior $R \sim L^{-2}$ also determines the dynamic critical exponent $z \approx 2$. Inset: The ratio $R(T=0.15)/R(T=0.30)$ versus $L$. It is exhibited
that the ratio shows downward curvature as $L$ is increased, which
implies that $R(T=0.15)$ vanishes in the thermodynamic limit
and that the resistive transition occurs at $T_c > 0.15$.
}
\label{fig:Rall}
\end{figure}

We plot in Fig.~\ref{fig:delta} for $L=6$ at $T=0.22$
(a) the time evolution of $L\Delta_y$ for 
a given disorder realization, and (b) $f(\Theta)$ obtained
from the average over 200 samples. The full line in Fig.~\ref{fig:delta}(b)
obtained from the fit determines the value of $R$.
Figure~\ref{fig:Rall} shows $R$ obtained in this way
as a function of $L$
at $T=0.30$, 0.25, 0.23, 0.22, 0.21, 0.20, and 0.15. 
For each $L$ and $T$, we divided the total number of samples, $N_s$, into five
groups, each of which contains $N_s/5$ samples, and $R$ was
calculated for each group, leading to the estimation of error bars
(standard deviations) in Fig.~\ref{fig:Rall}.
The linear resistance shows clear change of curvature
around $T \approx 0.22$: In the high-temperature regime $R$
appears to saturate as $L$ is increased, while in the
low-temperature regime $R$ drops down more rapidly as $L$
is increased. To see this behavior more clearly, we also
display in the inset of Fig.~\ref{fig:Rall} $R(T=0.15)/R(T=0.30)$
as a function of $L$ in log scales, which does not exhibit any
sign of saturation in terms of $L$: On the contrary, the downward
curvature implies that in the thermodynamic limit 
$R(T=0.15)$ vanishes.
These observations lead to the conclusion that
the 2D $XY$ gauge glass model has the {\it finite-temperature
resistive transition}. 

\begin{figure}
\centering{\resizebox*{!}{6.0cm}{\includegraphics{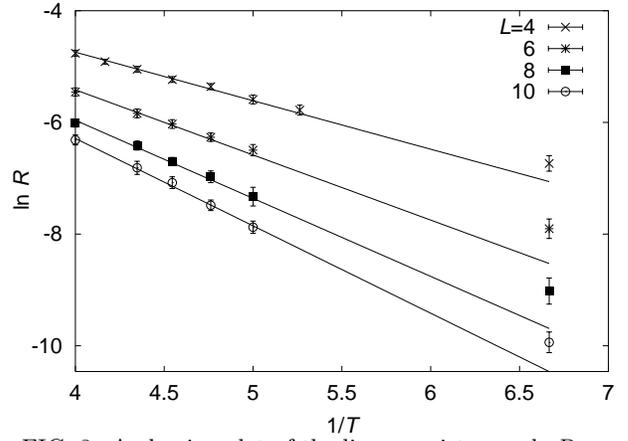}}}
\caption{Arrhenius plot of the linear resistance: $\ln R$ versus
$1/T$. $R$ shows a systematic deviation at the lowest temperature for 
all system sizes used.
}
\label{fig:arr}
\end{figure}

To obtain more precise value of $T_c$, we proceed to the
finite-size scaling analysis of the linear resistance.
In the dynamic scaling theory for the usual second-order 
phase transition in 2D,~\cite{ffh} the linear resistance has the
standard finite-size scaling form:
\begin{equation} \label{eq:scale}
R(L,T) = L^{-z} \rho\left( (T-T_c)L^{1/\nu} \right), 
\end{equation}
where $\rho(x)$ is the scaling function with the scaling variable $x$,
$z$ is the dynamic critical exponent, and the static critical
exponent $\nu$ is defined
by the divergence of the correlation length, i.e.,  $\xi \sim |T-T_c|^{-\nu}$.
At $T=T_c$, $R \sim  L^{-z}$ [see Eq.~(\ref{eq:scale})] 
and from Fig.~\ref{fig:Rall} we obtain $z \approx 2.0$. 
Although this value is much smaller than the values usually measured
in experiments ($z= 4 \sim 9$), 
it should be noted that the similar values
have also been observed in various numerical simulations with
RSJ dynamics and MC dynamics of 2D gauge glass model 
($z\approx 2.2$ in Ref.~\onlinecite{yhli} and 
$z \approx 2.4$ in Ref.~\onlinecite{mychoi}, respectively)
as well as the Langevin-type relaxational dynamics of vortices
in the presence of quenched impurities ($z\approx 2.1$, Ref.~\onlinecite{enomoto}).
Furthermore, a recent analytic calculation based on the
dynamic renormalization group method also has found 
$z=2$ for 2D gauge glass model with a purely relaxational
equations of motion.~\cite{gsjeon}
In contrast, widely believed zero-temperature 
transition in 2D gauge glass model implies that the linear
resistance should have Arrhenius form and thus corresponds
to $z = \infty$.  We also tried to fit our data in Fig.~\ref{fig:Rall} 
to the Arrhenius form~\cite{granato98,hyman} (see Fig.~\ref{fig:arr})
and found that the thermal activation barrier, which is proportional to
the slopes in Fig.~\ref{fig:arr}, strongly depends on the 
system size. Furthermore, $R$ is found to deviate from the Arrhenius form 
in a systematic way at the lowest temperature at all system sizes.

\begin{figure}
\centering{\resizebox*{!}{6.0cm}{\includegraphics{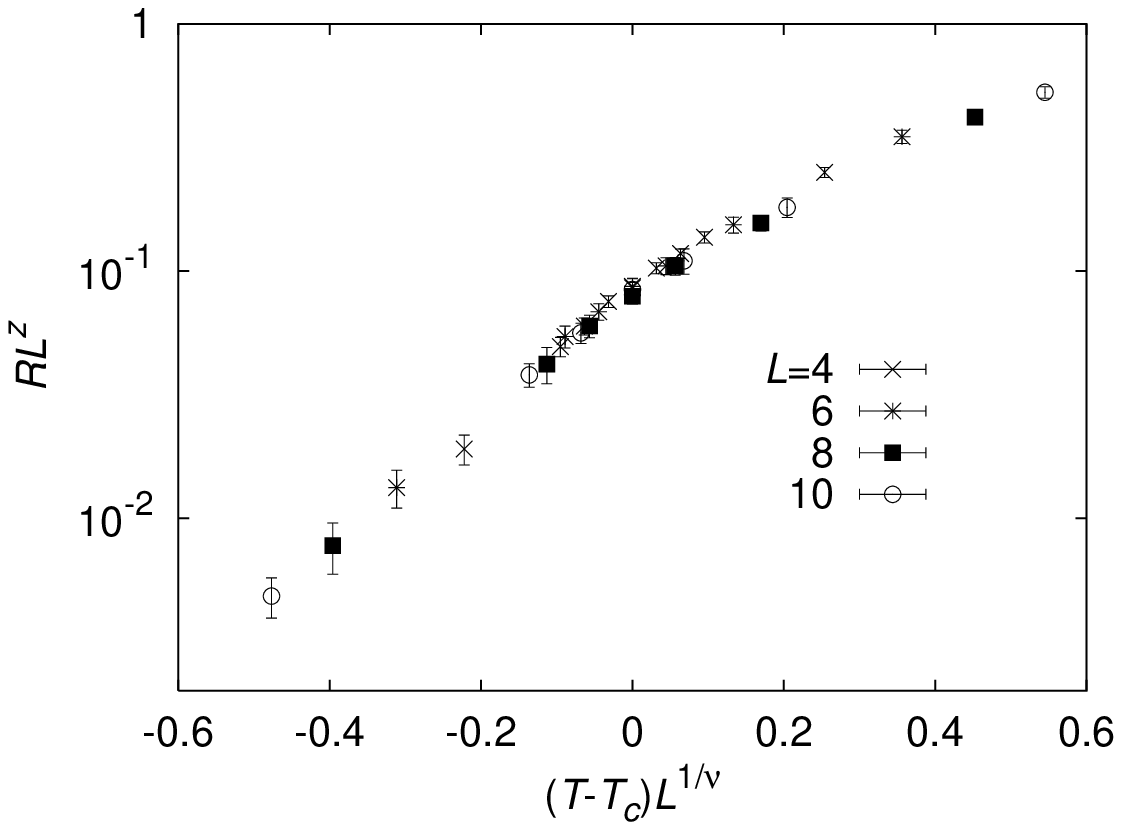}}}
\caption{Finite-size scaling plot of the linear resistance:
$R L^{z}$ versus the scaling variable $(T-T_c) L^{1/\nu}$ with
$T_c = 0.22$, $\nu = 1.2$, and $z=2.0$. (All data points in Fig.~\ref{fig:Rall}
have been used.)
}
\label{fig:scale}
\end{figure}

Figure~\ref{fig:scale} shows the scaling plot Eq.~(\ref{eq:scale})
with $z=2.0$, $T_c = 0.22$, and $\nu = 1.2$.
We tried to vary parameter values used in the scaling plot
and concluded
\begin{eqnarray}
T_c & = & 0.22(2), \\
z & = & 2.0(1), \\
\nu & = & 1.2(2) , 
\end{eqnarray}
where numbers in parenthesis denote errors in the last digits,
and $T_c \approx 0.22$ is in a good agreement with
Refs.~\onlinecite{mychoi} and \onlinecite{bjkim97}. It is worth mentioning
that the same method (calculation of the linear resistance
with FTBC accompanied by the finite-size scaling analysis)
has lead to a very precise determination of $T_c$ in 
3D $XY$ model.~\cite{lars} 
In recent studies of 2D $XY$ gauge glass model based on the same
RSJ dynamics,~\cite{granato98,hyman} the finite-size scaling 
analysis has not been used and the temperature range did not
cover the critical temperature ($T_c \approx 0.22$), 
leading to the different conclusion of the zero-temperature transition.

Recently, the possibility of the {\it quasi-long-range}
glass order with the vanishing glass order parameter~\cite{nishimori} in 2D gauge glass model
has been suggested in Ref.~\onlinecite{mychoi},
where it has been argued that the correlation length $\xi$ diverges
in the whole low-temperature phase.
Although our scaling plot in Fig.~\ref{fig:scale} was obtained
on the assumption of the usual second order transition, where
$\xi$ is finite both below and above $T_c$,  
it is still plausible to have the quasi-long-range glass order:
In this case, the quasi-criticality should be reflected in 
the temperature-dependent dynamic critical exponent 
below $T_c$ with $R(L,T) \sim L^{-z(T)}$,
while the high-temperature
phase still obeys the scaling form in Eq.~(\ref{eq:scale}).
Our current results cannot rule out this possibility,
and if we neglect the smallest size $L=4$ in Fig.~\ref{fig:Rall},
$R$ in the low-temperature phase indeed appears to exhibit the simple
algebraic form $R \sim L^{-z}$ with increasing $z$ as $T$ is decreased,
in accord with the idea of the quasi-long-range glass order.

In conclusion, we studied numerically the resistive transition
in the 2D $XY$ gauge glass model by using the RSJ dynamic equations
subject to the fluctuating twist boundary condition. The standard finite-size
scaling analysis applied to the linear resistances lead to the
strong evidence of the {\it finite-temperature} resistive transition
at $T_c = 0.22(2)$ with the dynamic critical exponent $z=2.0(1)$ and
the static exponent $\nu = 1.2(2)$. However, the nature of this
finite-temperature resistive transition needs to be
investigated more in detail.

\acknowledgments
The author is grateful to Prof. Petter Minnhagen for useful discussions.
This work was supported by the Swedish Natural Research Council
through contract FU 04040-332.

\newpage
\narrowtext

\vskip 1cm

\vskip -1cm

\vskip 1cm

\end{multicols}

\begin{thebibliography}{10}

\bibitem{mpafisher89}
M.~P.~A. Fisher, Phys. Rev. Lett. {\bf 62},  1415  (1989).

\bibitem{ffh}
D.~S. Fisher, M.~P.~A. Fisher, and D.~A. Huse, Phys. Rev. B {\bf 43},  130
  (1991).

\bibitem{YBCO}
R.~H. Koch, V. Foglietti, W.~J. Gallagher, G. Koren, A. Gupta, 
and M.~P.~A. Fisher, Phys. Rev. Lett. {\bf 63},  1511  (1989);
P.~L. Gammel, L.~F. Schneemeyer, and D.~J. Bishop, {\it ibid}. {\bf 66}, 953  (1991);
H.~K. Olsson, R.~H. Koch, W. Eidelloth, and R.~P. Robertazzi, {\it ibid}. {\bf 66},  2661  (1991);
C. Dekker, W. Eidelloth, and R.~H. Koch, {\it ibid}. {\bf 68},  3347 (1992);
P. {Voss-de Haan}, G. Jakob, and H. Adrian, Phys. Rev. B {\bf 60},  12443 (1999).

\bibitem{YBCO:2D}
C. Dekker, P.~J.~M. W{\"o}ltgens, R.~H. Koch, B.~W. Hussey, and A. Gupta,
Phys. Rev. Lett. {\bf 69},  2717  (1992);
P.~J.~M. W{\"o}ltgens, C. Dekker, R.~H. Koch, B.~W. Hussey, and A. Gupta,
Phys. Rev. B {\bf 52},  4536  (1995).

\bibitem{sawa}
A. Sawa, H. Yamasaki, Y. Mawatari, H. Obara, M. Umeda, and S. Kosaka,
Phys. Rev. B {\bf 58},  2868  (1998).


\bibitem{BiSCCO}
Q. Li, H.~J. Wiesmann, M. Suenaga, L. Motowidlow, and P. Haldar, 
Phys. Rev. B {\bf 50},  4256  (1994); Appl. Phys. Lett. {\bf 66},  637  (1995).

\bibitem{yamasaki}
H. Yamasaki, K. Endo, S. Kosaka, M. Umeda, S. Yoshida, and K. Kajimura,
Phys. Rev. B {\bf 50},  12959  (1994).

\bibitem{cieplak91}
M. Cieplak, J.~R. Banavar, and A. Khurana, J. Phys. A: Math. Gen. {\bf 24},
  L145  (1991).

\bibitem{Defect}
M.~J.~P. Gingras, Phys. Rev. B {\bf 45},  7547  (1992);
H.~S. Bokil and A.~P. Young, Phys. Rev. Lett. {\bf 74},  3021  (1995);
J.~M. Kosterlitz and M.~V. Simkin, {\it ibid}. {\bf 79},  1098  (1997);
J.~M. Kosterlitz and N. Akino, {\it ibid}. {\bf 81},  4672  (1998).

\bibitem{maucourt:moore}
M.~A. Moore and S. Murphy, Phys. Rev. B {\bf 50},  3450  (1994);
J. Maucourt and D.~R. Grempel, {\it ibid}. {\bf 58},  2654  (1998).

\bibitem{cieplak92}
M. Cieplak, J.~R. Banavar, M.~S. Li, and A. Khurana, Phys. Rev. B {\bf 45},
  786  (1992).

\bibitem{huse:reger}
D.~A. Huse and H.~S. Seung, Phys. Rev. B {\bf 42},  1059  (1990);
J.~D. Reger, T.~A. Tokuyasu, A.~P. Young, and M.~P.~A. Fisher, {\it ibid}.
  {\bf 44},  7147  (1991).

\bibitem{apyoung}
A.~P. Young,  in {\em Inaugural Symposium on Random Magnetism and High
  Temperature Superconductivity}, edited by W. Beyerman, N.~L. Huang-Liu, and
  D.~E. MacLaughlin (World Scientific, Singapore, 1994).

\bibitem{3DT}
K.~H. Lee and D. Stroud, Phys. Rev. B {\bf 44},  9780  (1991);
C. Wengel and A.~P. Young, {\it ibid}. {\bf 56},  5918  (1997);
T. Olson and A.~P. Young, cond-mat/9912291 (unpublished).

\bibitem{mpafisher91}
M.~P.~A. Fisher, T.~A. Tokuyasu, and A.~P. Young, Phys. Rev. Lett. {\bf 66},
  2931  (1991).

\bibitem{mychoi}
M.~Y. Choi and S.~Y. Park, Phys. Rev. B {\bf 60},  4070  (1999).

\bibitem{reger93}
J.~D. Reger and A.~P. Young, J. Phys. A: Math. Gen. {\bf 26},  L1067  (1993).

\bibitem{granato98}
E. Granato, Phys. Rev. B {\bf 58},  11161  (1998).

\bibitem{hyman}
R. A. Hyman, M. Wallin, M.~P.~A. Fisher, S.~M. Girvin, and A.~P. Young,
Phys. Rev. B {\bf 51},  15304  (1995).

\bibitem{yhli}
Y.-H. Li, Phys. Rev. Lett. {\bf 69},  1819  (1992).

\bibitem{bjkim97}
B.~J. Kim and M.~Y. Choi, Phys. Rev. B {\bf 56},  6007  (1997).

\bibitem{gsjeon}
G.~S. Jeon, K. Park, and M.~Y. Choi (unpublished).

\bibitem{enomoto}
Y. Enomoto and S. Maekawa, Physica C {\bf 274},  351  (1997).

\bibitem{bjkim99}
B.~J. Kim, P. Minnhagen, and P. Olsson, Phys. Rev. B {\bf 59},  11506  (1999);
B.~J. Kim and P. Minnhagen, {\it ibid}. {\bf 60}, 588 (1999).

\bibitem{eikman}
H. Eikmans and J.~E. van Himbergen, Phys. Rev. B {\bf 41},  8927  (1990);
J.~J.~V. Alvarez, D. Dom\'{\i}nguez, and C.~A. Balseiro, Phys. Rev. Lett. 
{\bf 79},  1373  (1997);
D. Dom\'{\i}nguez, {\it ibid}. {\bf 82},  181  (1999).

\bibitem{weber}
H. Weber and H.~J. Jensen, Phys. Rev. Lett. {\bf 78},  2620  (1997).

\bibitem{olsson}
P. Olsson, Phys. Rev. B {\bf 46}, 14598  (1992); {\bf 52},  4511  (1995); 4526  (1995).

\bibitem{lars}
(a) L.~M. Jensen, B.~J. Kim, and P. Minnhagen, Europhys. Lett. {\bf 49}, 644 (2000); (b) Phys. Rev. B (in press).

\bibitem{batrouni85}
G.~G. Batrouni, G.~R. Katz, A.~S. Kronfeld, G.~P. Lepage, B. Svetitsky,
and K.~G. Wilson, Phys. Rev. D {\bf 32},  2736  (1985).

\bibitem{reif}
F. Reif, {\em Fundamentals of statistical and thermal physics} (McGraw-Hill,
  New York, 1965).

\bibitem{foot:R}
For the details of the approximation made see Ref.~\onlinecite{lars}(b). The
  similar approximation has also been made in Ref.~\onlinecite{granato98} for the
  system with open boundary.


\bibitem{nishimori}
It has been shown analytically that in 2D $XY$ gauge glass model
the glass order parameter vanishes at any
finite temperatures in H. Nishimori, Physica A {\bf 205},  1  (1994).


\end{thebibliography}
\end{document}